\documentclass{article}

\usepackage[preprint]{neurips_2025}

\usepackage[hidelinks]{hyperref}

\usepackage[utf8]{inputenc} %
\usepackage[T1]{fontenc}    %
\usepackage{hyperref}       %
\usepackage{url}            %
\usepackage{booktabs}       %
\usepackage{amsfonts}       %
\usepackage{nicefrac}       %
\usepackage{microtype}      %
\usepackage{xcolor}         %
\usepackage{algorithmic,algorithm,float}

\usepackage{our_style}
\usepackage{graphicx}
\usepackage{wrapfig}

\title{Learning From the Past with Cascading Eligibility Traces}

\usepackage{authblk}

\newcommand{\equal}{\textsuperscript{*}} 
\newcommand{\supervision}{\textsuperscript{$\dagger$}} 

\author[1,2]{Tokiniaina Raharison Ralambomihanta\equal}
\author[1,3]{Ivan Anokhin\equal}
\author[1,4,5]{Roman Pogodin\equal}
\author[1,6,7]{Samira~Ebrahimi~Kahou\supervision}
\author[1,8]{Jonathan Cornford\supervision}
\author[1,4,5,9,10]{Blake A. Richards\supervision}

\affil[1]{Mila -- Quebec Artificial Intelligence Institute}
\affil[2]{Department of Bioengineering, McGill University}
\affil[3]{Université de Montréal}
\affil[4]{Department of Neurology \& Neurosurgery, McGill University}
\affil[5]{Montreal Neurological Institute, McGill University}
\affil[6]{University of Calgary}
\affil[7]{CIFAR AI Chair}
\affil[8]{School of Computer Science, University of Leeds}
\affil[9]{School of Computer Science, McGill University}
\affil[10]{CIFAR Learning in Machines and Brains Program}

\affil[*]{Equal contribution}
\affil[$\dagger$]{Co-senior authors}

\affil[ ]{{\tt\{toky.raharison-ralambomihanta, ivan.anokhin,
roman.pogodin, ebrahims, cornforj, blake.richards\}@mila.quebec}}

\begin{document}

\maketitle

\begin{abstract}
  Animals often receive information about errors and rewards after a significant delay. For example, there is typically a delay of tens to hundreds of milliseconds between motor actions and visual feedback. The standard approach to handling delays in models of synaptic plasticity is to use eligibility traces. However, standard eligibility traces that decay exponentially mix together any events that happen during the delay, presenting a problem for any credit assignment signal that occurs with a significant delay. Here, we show that eligibility traces formed by a state-space model, inspired by a cascade of biochemical reactions, can provide a temporally precise memory for handling credit assignment at arbitrary delays. We demonstrate that these cascading eligibility traces (CETs) work for credit assignment at behavioral time-scales, ranging from seconds to minutes. As well, we can use CETs to handle extremely slow retrograde signals, as have been found in retrograde axonal signaling. These results demonstrate that CETs can provide an excellent basis for modeling synaptic plasticity.
\end{abstract}

\section{Introduction}

Learning requires a mechanism for assigning credit for errors and successes to past neural activity \citep{gerstner2018eligibility}. In biological learning, the signals necessary for this credit assignment often arrive after a long delay: for example visual feedback typically follows motor actions with a delay of tens to hundreds of milliseconds \citep{omrani2016distributed, scott2016functional}. As well, candidate biological signals for communicating credit, such as retrograde axonal molecular messengers, can take minutes to propagate through a neural circuit \citep{oztas2003neuronal, fitzsimonds1998retrograde}. Given these delays, biological learning necessarily requires mechanisms for bridging temporal gaps ranging from hundreds of milliseconds to several minutes. 

At the level of synaptic weight changes, the traditional solution to handling delays is to use exponentially decaying synaptic eligibility traces \citep{gerstner2018eligibility, shouval2025eligibility}, which are decaying records of synaptic activity.
When the neural activity and the corresponding credit assignment signal are separated by few intervening events, such delays will have minimal impact on learning. 
However, in general, ongoing neural activity will override past activity relevant to the current reward or error signal. 
Therefore, traditional eligibility traces are not well-suited for the temporal scale of credit assignment delays in biological learning. 

Here, we present a generalization of traditional eligibility traces. Inspired by synaptic biochemical cascades \citep{zhang2021quantitative}, we model eligibility traces as state-space models that incorporate a cascade of synaptic memory traces. These cascading eligibility traces (CETs) provide a more precise temporal window for credit assignment, but are also consistent with biological mechanisms of synaptic plasticity \citep{friedrich2011spatio, fusi2005cascade, zhang2021quantitative}.

We present a series of results that demonstrate the utility of CETs for credit assignment with biologically realistic delays. Specifically, we show that we can engage in both supervised and reinforcement learning in multi-layer networks under two distinct delay scenarios. First, we examine learning situations where the delays are consistent across layers of the network, as would be the case for various models of biological credit assignment in which a learning signal is broadcast across layers (e.g. direct feedback alignment \citep{nokland2016direct} and other local learning rules \citep{fremaux2016neuromodulated,ororbia2023brain}). Second, we show that CETs work when delays are stacked through the network, such that late layers receive credit signals sooner than early layers. We provide evidence that this approach works in both scenarios under a variety of biologically relevant delays, ranging from hundreds of milliseconds to minutes. Notably, the fact that CETs work when delays are stacked across layers and last for minutes shows that CETs could be applicable to credit assignment signals carried by retrograde axonal signals or neuropeptides, making this approach relevant for a number of biologically plausible credit assignment models \citep{liu2022biologically, fan2024contribute}.

Altogether, our results indicate that CETs are a promising approach for handling delayed credit assignment signals in models of biological learning. More broadly, this provides a general framework for reasoning about synaptic memory in real neural networks. Our code is available at: \url{https://github.com/avecplezir/CET}.

\section{Related work}

Our work is related to and builds upon several strands of research on synaptic plasticity and biological credit assignment.

Eligibility traces (ETs) have long been a dominant framework for modeling how synaptic plasticity mechanisms may bridge temporal gaps between neural activity and feedback \citep{gerstner2018eligibility,shouval2025eligibility}. Related theoretical extensions include ETs to approximate backpropagation through time (BPTT; \cite{bellec2020solution}). Experimental evidence for ETs is well-established, with multiple studies reporting how synaptic changes can be induced by reward signals arriving seconds to minutes after neural activity \citep{brzosko2015retroactive,he2015distinct,bittner2017behavioral,suvrathan2019beyond}. 

In some experimental results, the timing of maximum synaptic change is tuned to specific delays (e.g. 120ms in the cerebellum \citep{suvrathan2016timing} and 2s in the striatum \citep{shindou2019silent}). 
This is in contrast to traditional ETs, and indicates a plasticity rule that encodes temporal structure in addition to the presence of past activity. One approach to model these findings is to combine two independent ETs for potentiation and depression to produce a composite ET that peaks at a required time delay \citep{he2015distinct,huertas2016role}. 
This approach is conceptually similar to our CET model with 2 states.
However, as we discuss below, this approach is restricted to producing ETs with a broad integration window, making it suitable for short delays only. As we show in our work, CETs with a larger number of states overcome this issue.

An example of extreme delays in plasticity-related signals is retrograde axonal signaling: i.e. ``backward'' propagation of chemical signal through the axon and synapses \citep{maday2014axonal, alger2002retrograde, fitzsimonds1998retrograde}. 
These signals play a role in activity-dependent synaptic plasticity at the level of individual synapses \citep{regehr2009activity}, and have been suggested to coordinate plasticity across several neurons \citep{fitzsimonds1997propagation, hui2000selective, du2004rapid}.

However, retrograde signals have generally been discarded as a component of credit assignment (e.g. \cite{lillicrap2016random}) because retrograde axonal signaling is extremely slow (on average 1.31$\mu$m/s; \cite{cui2007one}), meaning that any error signal delivered via retrograde signaling would arrive minutes after the relevant neural activity. Nevertheless, there's been recent interest in this approach \citep{fan2024contribute}. Here, we study to what extent delays on the order of retrograde process timescales could be compensated with CETs.

In parallel to neuroscience, the deep learning community has also worked on the the problem of delayed feedback from the perspective of decoupling the forward and backward passes for efficiency \citep{jaderberg2017decoupled,malinowski2020sideways}.
More generally, there is an extensive body of related work modeling how neural circuits may estimate and communicate credit in a biological plausible manner \citep{lillicrap2020backpropagation}. This includes credit computations with dendrites and bursts \citep{greedy2022single, payeur2021burst, sacramento2017dendritic}, and neuropeptides \citep{liu2022biologically}.

\section{Cascading eligibility traces (CETs)}\label{sec:cet}
Synaptic plasticity for learning always requires some memory for presynaptic activity in the network. Consider a network containing a neuron with activity $z_t = f(\x_t\T\w)$, where $\w$ are the synaptic weights and $\x_t$ are the presynaptic inputs to the neuron at time $t$. To minimize a loss $L$, synaptic changes in $\w$ can follow the negative gradient over the loss $L$,
\begin{equation}
    -\parderiv{L(\x_t\T\w)}{\w} = -\parderiv{L(z_t)}{z_t}\,f'(\x_t\T\w)\,\x_t \equiv -\delta_t\,f'(\x_t\T\w)\,\x_t\,.
    \label{eq:neuron_grad}
\end{equation}

This gradient-based formulation of plasticity \cref{eq:neuron_grad} covers various forms of biological learning. For example, Hebbian learning can be recovered by using the loss $L(z_t)=-z_t^2$ \, and layer-wise learning rules can be defined similarly. And, of course, backpropagation follows the same chain rule logic for the loss defined over several layers of neurons. 

\begin{figure}[t]
    \centering
    \includegraphics[width=\linewidth]{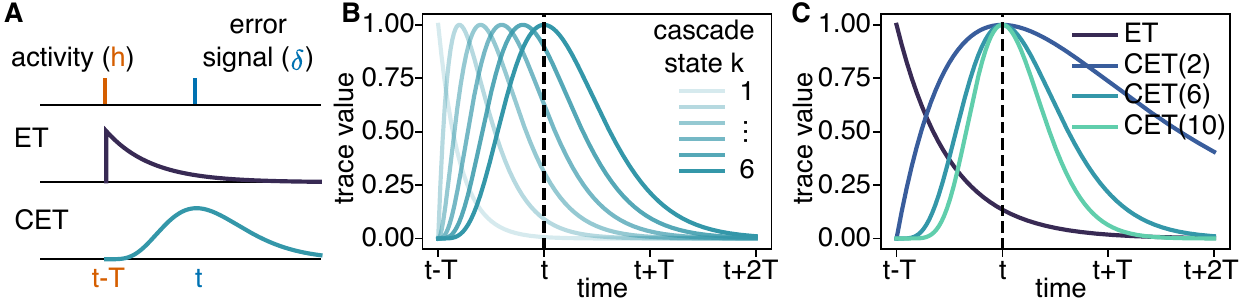}
    \caption{\textbf{A.} Learning with eligibility traces: neural activity $h$ is followed by a delayed error signal $\delta$. The standard eligibility trace (ET) is an exponentially decaying trace of $h$ that can be matched to the error signal at time $t$. The cascading ET (CET) reflects $h$ like a regular ET, but peaks at the required time $t$. 
    \textbf{B.} Time evolution of each state of a 6-state CET with a delay $T$ and a unit input at $t-T$.
    \textbf{C.} Comparison of a standard ET and CETs with 2/6/10 states representing delay $T$ for a unit input at $t-T$.}
    \label{fig:cet_diagrams}
\end{figure}

Importantly, these updates require that the credit assignment signal $\delta_t$ is paired with the appropriate presynaptic inputs, $\x_t$. Hence, in the presence of any delays in computation of the credit assignment signal a learning system would face a temporal mismatch problem: if it takes $T$ seconds to calculate and propagate the credit assignment, then at time $t$ the error signal $\delta_t$ received by a neuron would have to be matched to an older presynaptic activity memory $\x_{t-T}$ (see \cref{fig:cet_diagrams}A, top row). If learning is done in phases this need not be problematic. But, if learning happens online, as is likely the case in real brains, neural activity would correspond to the current time point, $\x_{t}$ only, so the previous presynaptic activity information, $\x_{t-T}$, would have to be somehow stored by the synapses. 

Eligibility traces (ET) represent the classic solution to this problem: they add a memory component to the synapse that keeps track of recent activation for a single presynaptic neuron. Here, we will pick one weight $w^i$ and the corresponding input $x_t^i$, and discuss an ET $h^{\mathrm{ET}}_{t}$ such that changes in $w^i$ are proportional to $-\delta_{t}\,h^{\mathrm{ET}}_{t}$ (as in \cref{eq:neuron_grad}; dropping the index $i$ from $h^{\mathrm{ET}}_{t}$ for convenience).

Denoting the Hebbian-like term $h=f'(\x_t\T\w)\,x_t^i$ (in the sense of it being a product of pre- and postsynaptic factors, $x_t^i$ and $f'(\x_t\T\w)$ correspondingly),
\begin{equation}
    h^{\mathrm{ET}}_t = \int_{0}^t e^{-\gamma\, (t-s)}\,h_s\, ds\,,
    \label{eq:et_def}
\end{equation}
where $\gamma > 0$ is a discount factor. The main advantage of ETs is that they're easy to implement:
\begin{equation*}
    \dot h^{\mathrm{ET}}_t = -\gamma\, h^{\mathrm{ET}}_t + h_t\,.
    \label{eq:et_def_ode}
\end{equation*}

ETs effectively convolve the presynaptic activity $h_s$ with an exponential kernel $g(t)$, i.e.
\begin{equation}
    h^{\mathrm{ET}}_t = (g * h)(t) = \int_{0}^t g(t-s)\,h_s\, ds\,,
    \label{eq:et_conv}
\end{equation}

and use this as a means of weighting past activity for combining it with credit assignment signals. One of the appeals of ETs as a solution to delayed credit assignment signals is that they do not require extensive memory, and are therefore a biologically plausible approach for learning.

Notably, the classic form of ET assigns the maximal trace values to the most recent time-points $s=t$ in \cref{eq:et_conv}. That is appropriate for situations in which there are few intervening presynaptic events between times $t$, when the credit assignment signal arrives, and $t-T$, when the presynaptic activity occurred (as in \cref{fig:cet_diagrams}A). But, if $h_t$ changes frequently relative to the delay in the credit assignment signal then gradients calculated with classic ETs, i.e. $\delta_{t}\,h^{\mathrm{ET}}_{t}$, can be a poor approximation of the true gradient, $\delta_{t}\,h_{t-T}$. 

Ideally, when we consider the long delays faced by biological learning agents we would have ETs satisfying two conditions. First, the maximal value of the synaptic trace should occur at a delay of $s=t-T$, rather than $s=t$ in \cref{eq:et_conv}. Second, it is better if we can use a more precise weighting of the past, i.e. if the ET values $g(t-s)$ are as small as possible for anything other than $s=t-T$. 

Classic ETs do not provide these characteristics. Combined LTP \& LTD eligibility traces \citep{he2015distinct,huertas2016role} satisfy the first condition, but not the second. They effectively take a difference of two standard ETs in \cref{eq:et_def} to convolve past activity with $g(t)=\exp(-\gamma_{\mathrm{P}} t)-\exp(-\gamma_{\mathrm{D}} t)$ (as in \cref{eq:et_conv}). While this $g$ peaks with a delay $T=(\log\,\gamma_{\mathrm{D}}-\log\,\gamma_{\mathrm{P}}) / (\gamma_{\mathrm{D}}-\gamma_{\mathrm{P}})$, it keeps a large weight for more recent points.

\begin{wrapfigure}{r}{0.50924369747 \textwidth}
\vspace{-0.5cm}
\includegraphics[width=\linewidth]{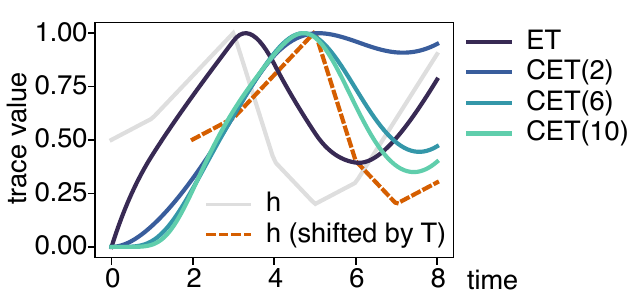} 
\caption{Representation of the input signal $h$ (gray) with a $T=2$ delay (dashed orange line) using a standard eligibility trace (ET) and cascading ETs (CETs) of different orders. Only CETs of higher order reflect the time evolution of the input (i.e. match the orange line).}
\label{fig:cet_integration}
\vspace{-2.5cm}
\end{wrapfigure}
As a flexible solution to both of these problems, we propose using ETs constructed from a simple state-space model:
\begin{equation}
\begin{split}
    \dot h^{1}_t \,&= -\alpha\, h^{1}_t + h_t\,,\\
    &\dots\\
    \dot h^{k}_t \,&= -\alpha\, h^{k}_t + h_t^{k-1}\,,\\
    &\dots\\
    \dot h^{\mathrm{CET}}_t \,&= -\alpha\, h^{\mathrm{CET}}_t + h_t^{n-1}\,.\\
\end{split}
\label{eq:ssm}
\end{equation}

\Cref{eq:ssm} can be used as a model of a cascade of biochemical reactions. This could involve, for example, a cascade of phosphorylation processes or enzymatic reactions \citep{zhang2021quantitative}.

This form of CET gives us the following formulation (see \cref{sec:et_derivations} for a derivation):
\begin{equation}
    h^{\mathrm{CET}}_t = \frac{1}{(n-1)!}\int_{0}^t (t-s)^{n-1}e^{-\alpha\,(t-s)}\,h_s\, ds\,,
    \label{eq:ssm_et_def}
\end{equation}
which for $\alpha=\frac{n-1}{T}$ convolves the presynaptic activity with a kernel $g(t)\propto t^{n-1}e^{-\alpha\,t}$ that peaks at $t=T$ (\cref{fig:cet_diagrams}C). 
For classical ETs (which correspond to a single-state model of $n=1$) we either set the decay to be $\alpha=\frac{1}{T}$ (for supervised learning) or we conduct a grid search on this hyperparameter (for reinforcement learning).  

The dynamical system in \cref{eq:ssm} is defined by two parameters: the number of states $n$ and the decay term $\alpha$. Increasing $n$ while keeping the peak-time fixed leads to a narrower kernel (see \cref{fig:cet_diagrams}C for a visualization of different kernels $g(t)$), but having even two states (instead of one for standard ETs) can account for delayed signals. However, to accurately represent delayed signals, more states are typically needed (see \cref{fig:cet_integration}). Importantly, the position of the peak is fully determined by a scalar parameter $\alpha$. As a general model of activity-dependent plasticity, CETs can thus adapt to different brain areas and error computation mechanisms with varying delays \citep{suvrathan2016timing,shindou2019silent}. We now demonstrate that CETs can be used for delayed credit assignment in a variety of conditions.

\section{Experiments}\label{sec:experiments}

We illustrate the influence of CETs on learning with delays in two scenarios\footnote{Code is available at: \url{https://github.com/avecplezir/CET}}: (1) learning with delays on behaviorally relevant timescales (e.g. on the order of seconds) in \cref{subseq:btsp}; (2) credit propagation through very slow chemical signals (e.g. retrograde axonal signaling \citep{fitzsimonds1997propagation}) in \cref{subseq:retro}. In our simulations we assume that each input lasts for $200$ ms, which is roughly one saccade or one theta cycle in the brain \citep{young1963variable}. Thus, a single time-step in the simulation is treated as a $200$ ms, so a delay of $T=1$ s would mean that the $\delta$ signal arrives 5 time-steps after the input is initially presented to the network. Put another way, with a simulated delay of $T=1$ s there are $4$ image presentations that occur after the initial image presentation and before the $\delta$ signal for that image arrives.

In (1), we assume that the error signal $\delta$ is propagated to all layers simultaneously since credit signals propagated via action potentials could be transmitted to the entire network in parallel. As well, we calculate the error signal explicitly, but we note that this calculation could easily be substituted with any of the available mechanisms for biologically plausible error calculation (e.g. 3-factor Hebbian learning rules \cite{fremaux2016neuromodulated}). 

In (2), we consider a much longer delay $T$ of 2 minutes, which corresponds to roughly the amount of time it takes for chemical signals to travel backwards along the axon. At the speed of $1.31\,\mu$m/s \citep{cui2007one}, this covers roughly $\sim160\,\mu$m, corresponding to the typical $<200\,\mu$m distance in the cortex \citep{song2005highly,cui2007one}. As well, in-line with propagation of a retrograde signal, we assume that the delays stack up over layers. Thus, the last layer has no delay, the penultimate layer has a delay of $T=2$ minutes, the next layer has a delay of $T=4$ minutes, and so on. Thus, each preceding layer's delay is increased by $T$. 

We use two types of tasks: supervised image recognition on MNIST \citep{lecun1998mnist} and CIFAR-10 \citep{krizhevsky2014cifar}, and reinforcement learning on state-based environments (namely CartPole and LunarLander), as well as on a more complex visual environment (namely MinAtar/SpaceInvaders \citep{young2019minatar}, which use raw pixel observations as input).
We use a 3-layer MLP (\textit{input} $\rightarrow$ 512 $\rightarrow$ 512 $\rightarrow$ 10) for MNIST and a small CNN with 3 convolutional layers (\textit{input} $\rightarrow$ 32 $\rightarrow$ 64 $\rightarrow$ 128) and two linear layers (512 $\rightarrow$ 10) for CIFAR-10. 

For RL, we use a 3-layer MLP with a hidden dimension of 256 that we train with the Actor-Critic method, and report results over 3 seeds. To simplify training in the delayed setup, only the Actor is trained with a delayed error signal, while the Critic is updated via standard backpropagation. The Actor is trained using an online implementation of the $\lambda$-return via \textit{RL eligibility traces} \citep{sutton2018eligibility}.

Other experimental details (hyperparameters, compute resources) can be found in \cref{app:sec:experiments}. The PyTorch \citep{paszke2019pytorchimperativestylehighperformance} implementation and experiments are provided in the Supplementary Material.

\subsection{Learning with delays on behaviorally relevant timescales}\label{subseq:btsp}

\begin{figure}
    \centering
    \includegraphics[width=\linewidth]{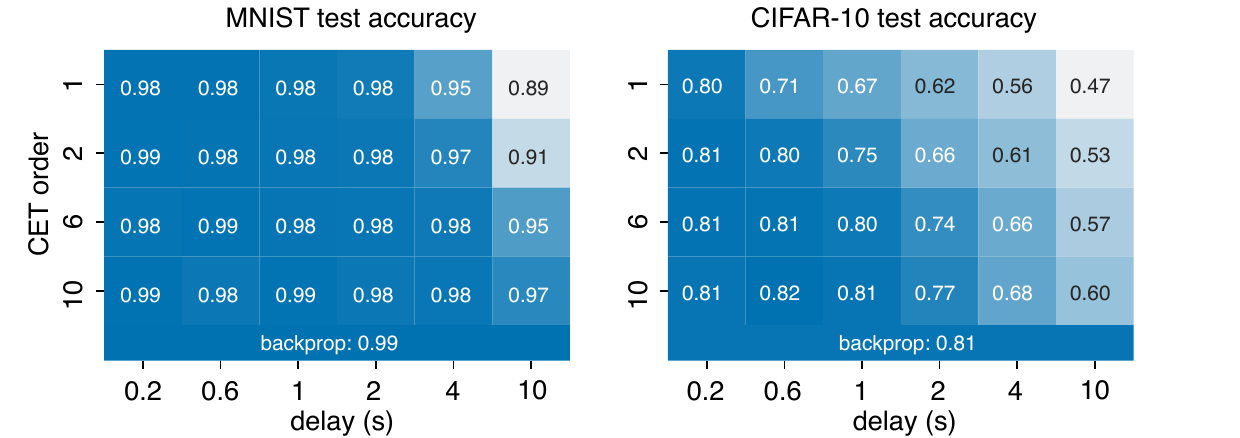}
    \caption{Accuracy for MNIST and CIFAR-10 datasets across varying numbers of CET states and delays on behaviorally relevant timescales. A single state (top row) corresponds to standard ETs.}
    \label{fig:btsp_visual}
\end{figure}

On MNIST, we observed that classical ETs (corresponding to a CET with one state) maintain strong performance up to delays of two seconds, i.e. up to 10 image presentations before a $\delta$ arrives (\cref{fig:btsp_visual}, left). This shows that classical ETs can remain effective for relatively simple tasks and short delays. 
However, their performance breaks down at longer delays of $T\geq 4$ s. At these longer delays we can see that increasing the number of states in the CETs improves performance, and can keep the accuracy level high at up to $10$ s delay (50 image presentations). Past this point, we found that only a perfect eligibility trace (i.e. an infinite number of states corresponding to a Dirac delta memory) would preserve performance. 

The results with CIFAR-10 were even more pronounced (\cref{fig:btsp_visual}, right). Classical ET performance deteriorates at any delay tested and rapidly decreases, showing that more complex visual tasks are less robust to imprecise time resolution. 
A key observation is the gradient in performance, with accuracy generally improving with the number of states in the CETs and decreasing with delays. 
This trend reflects how the CET impulse response becomes increasingly concentrated around the target delay as the number of states increases, which provides finer temporal resolution. %

\begin{figure}[h!]
    \centering
\includegraphics[width=\linewidth]{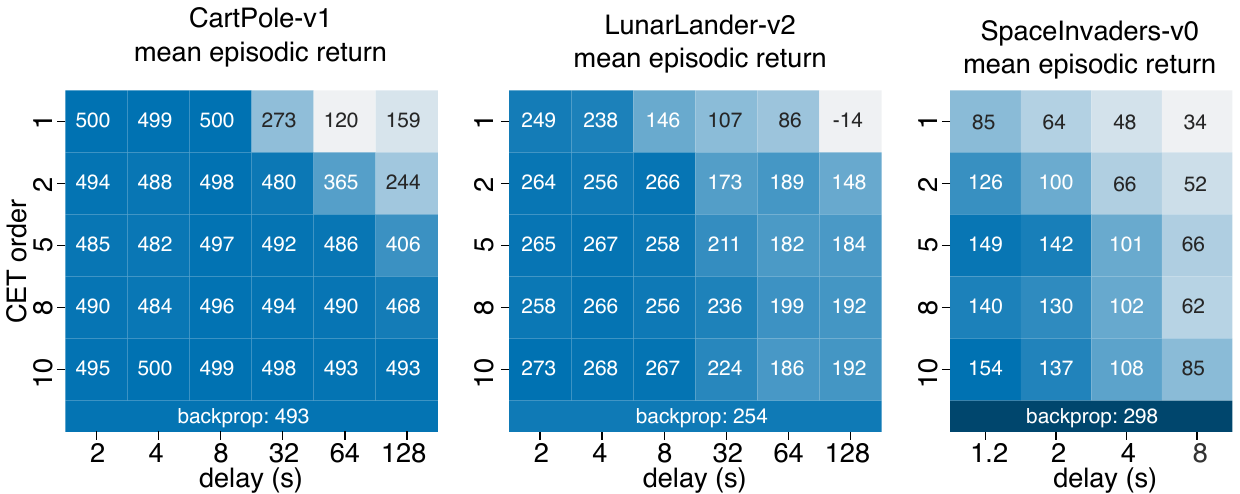}
\caption{Mean episodic return for different RL environments across varying numbers of CET states and delays on behaviorally relevant timescales. A single state (top row) corresponds to standard ETs. 
}
\label{fig:btsp_rl}
\end{figure}

We observe the same trend for the RL tasks (see \cref{fig:btsp_rl}): shorter delays and a higher number of CET states result in better performance. 
Note that CartPole and LunarLander are simple RL tasks and remain solvable even with long delays and fewer states in the CETs. 
In contrast, MinAtar/SpaceInvaders (\cref{fig:btsp_rl}, right) is a more complex, image-based environment where performance begins to degrade more quickly when a delay is introduced. In fact, CET performance is at best only half the performance of a perfect memory, even at the shortest delay we tested. 
We therefore hypothesize that precise credit assignment, without mixing nearby time points, is especially important for complex,  non-i.i.d. tasks. Altogether, our results demonstrate that at behaviorally relevant time delays higher-order CETs can greatly enhance performance beyond that achieved by classical ETs, particularly at long delays. However, they cannot fully compensate for delays in highly complex, non-i.i.d. tasks.

\begin{figure}[h!]
    \centering
    \includegraphics[width=\linewidth]{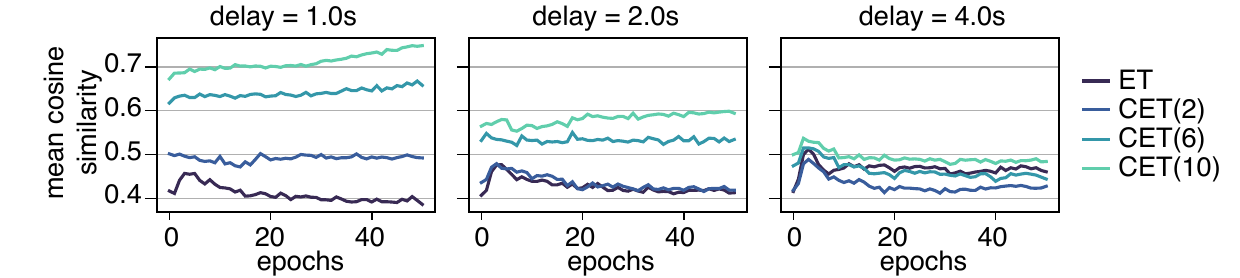}
    \caption{Average cosine similarity over all layers between true gradients and gradients computed with either ETs or CETs for the CIFAR-10 dataset.}
    \label{fig:btsp_sim}
\end{figure}

To better understand the reasons for the performance we examined how well the weight updates were aligned to the true gradient. We measured the cosine similarity, $\bb a\T\bb b / (\|\bb a\|\,\|\bb b\|)$, between the vector of weight updates given by our CETs, $\bb a$, and the true gradient as calculated by backpropagation, $\bb b$. In \cref{fig:btsp_sim}, we plot cosine similarity for all CET models for delays of 1, 2, and 4 s on CIFAR-10 (as performance differences were noticeable across these delays in \cref{fig:btsp_visual}).
We observed that at all times during training, and at shorter delays, an increase in the number of states in the CETs lead to better alignment with the true gradient (\cref{fig:btsp_sim}, left and center). However, as the delay increases, the alignment drops even for higher-order CETs (e.g. with 10 states; \cref{fig:btsp_sim}, right). When we broke this down by layer, we observed similar patterns (\cref{app:sec:supp_results}).

\subsection{Computation with extremely long delays for retrograde axonal signaling}\label{subseq:retro}

We next investigated the possibility of using CETs to model situations with very long, and accumulating, delays. Here, the goal was to consider delays introduced by chemical signals (e.g. retrograde axonal signaling) which could in principle be used for credit assignment \citep{fan2024contribute}, but would take minutes to propagate from synapses back to cell bodies. 
We assume that we do not have to solve the weight transport problem of backpropagation \citep{grossberg1987competitive}, since retrograde signals could easily have access to synaptic weight values \citep{fan2024contribute}. (See Discussion.)
However, the approach with CETs here could also easily be used in conjunction with other solutions to weight transport, including feedback alignment \citep{lillicrap2016random} or feedback learning mechanisms \citep{akrout2019deep}.
Additionally, we assume that the calculation of the $\delta$ signals has access to the post-synaptic activation derivative $f'(\x\T\w)$ at the appropriate delay, which implies another memory mechanism at the soma, rather than the synapse. 
This could be modeled with CETs as well, but we leave that for future work.

\begin{wrapfigure}{r}{0.3294117647 \textwidth}
\includegraphics[width=\linewidth]{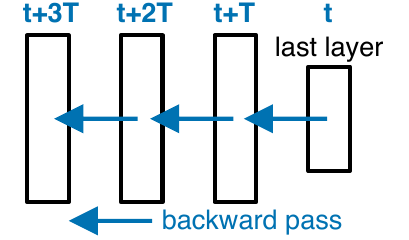} 
\caption{Backprop using retrograde axonal signaling results in delay accumulation: in the last layer, the forward and backward signal are computed simultaneously at time $t$. Each consecutive layer of the backwards pass takes $T$ more seconds (time taken by retrograde signaling).}
\label{fig:retro_diagram}
\vspace{-0.5cm}
\end{wrapfigure}

An additional consideration that we took into account here is that if credit signals were propagating backwards via retroaxonal biochemical transmission, then error signals would
take progressively more time as they travel across a number of synaptic steps (i.e. network depth).
Therefore, in a feedforward network if we assume that a single layer takes $T$ time to propagate the error signal backwards, then a layer $m$ synaptic steps back will receive the error signal at time $t=(m-1)T$ (\cref{fig:retro_diagram}). 

Finally, to handle very long delays on visual tasks (MNIST, CIFAR-10), we reduced the number of inputs being stored in CETs. 
To achieve this, inspired by work on reducing the energetic costs of plasticity \citep{van2024competitive}, we assume that the CETs are modulated by an additional ``salience signal'' that zeros out the input to the CETs unless the loss is very large. We use 1.25 \% of points with the largest losses in a batch (keeping their positions in the batch, such that position index encodes time; see \cref{sec:mat_conv} for details). 
We assume that salience signal is computed with little delay and can be propagated to the whole network (e.g. via neuromodulators or plateau potential in apical dendrites \citep{sacramento2017dendritic, dabney2020distributional}). 
To handle long retrograde signal delays in RL without sparsification, we simplify the setting by assuming each time step lasts $300$ ms instead of $200$ ms as in other our experiments. This corresponds to a delay of $400$ frames for the second layer and $800$ frames for the first layer.

\begin{figure}[h!]
    \centering
    \includegraphics[width=\linewidth]{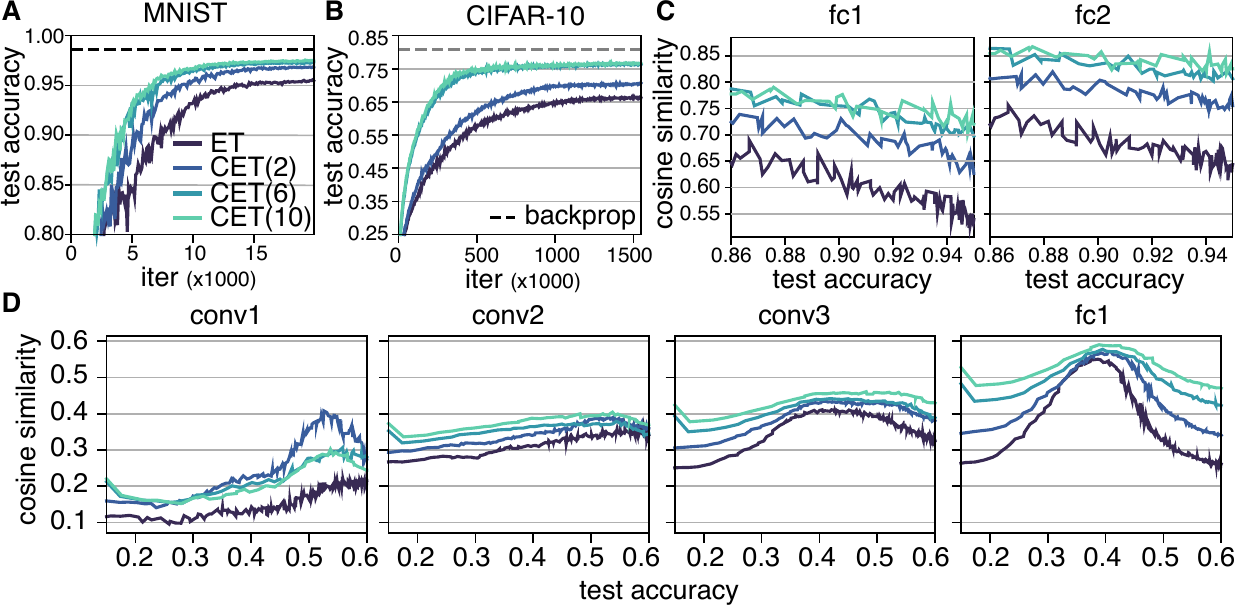}
    \caption{\textbf{A.} Test accuracy on MNIST as a function of number of CET states for the retrograde experiments. \textbf{B.} Same as \textbf{A}, but on CIFAR-10. \textbf{C.} Cosine similarity between the true gradient and the ET/CET approximation across different test accuracies (see \textbf{A}). Each plot shows an individual layer of an MLP during training on MNIST. \textbf{D.} Same as \textbf{C}, but each plot shows an individual layer of a CNN during training on CIFAR-10.}
    \label{fig:retro_visual}
\end{figure}

When training networks on visual tasks with large stacking delays across layers, we observed that the performance increased with increasing CET order (\cref{fig:retro_visual}). 
Moreover, networks with different CET orders trained at markedly different rates, with higher orders learning faster (\cref{fig:retro_visual}) . 
This was more pronounced for the deeper convolution network trained on CIFAR-10 (\cref{fig:retro_visual}B) than the shallow MLP trained on MNIST (\cref{fig:retro_visual}A). 
Given that the last layer was trained without delay in these experiments, these results must be due to the impact of delays on learning in the intermediate representations. 
 
To understand the performance differences for different orders of CET we analyzed gradient alignment during training. Because networks learned at different speeds, which affects the dynamics of gradient alignment (\cref{app:sec:supp_results}), we used test accuracy as the independent variable, rather than training iteration.
On both MNIST and CIFAR-10 we observed that gradient alignment degrades for the earlier layers, as is expected for the increasing delays (\cref{fig:retro_visual}C,D).
Across all layers, increasing CET order was associated with an increase in gradient alignment, in line with task performance.
However, higher orders were unable to fully recover alignment with the gradient.

\begin{figure}[ht]
    \centering
    \includegraphics[width=\linewidth]{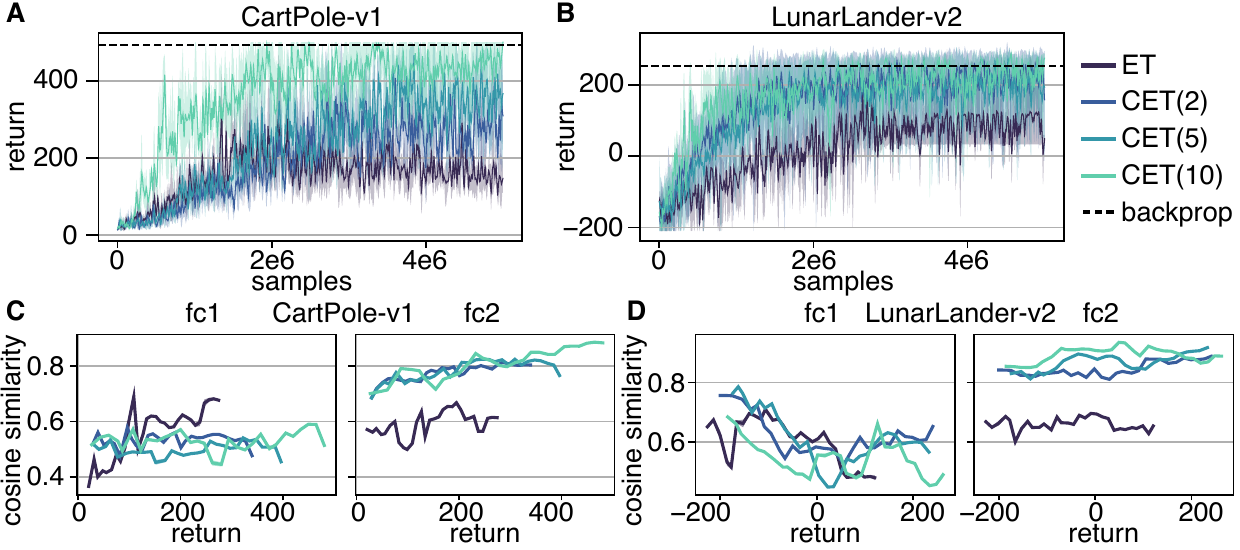}
    \caption{\textbf{A.} Episodic return on CartPole-v1 during training in the retrograde experiments. Solid lines: mean (3 seeds); shaded area: min/max values; dashed line: mean final backprop performance. \textbf{B.} Same as \textbf{A} but for LunarLander-v2. \textbf{C.}  Mean cosine similarity (over cosine similarity values assigned to binned return values) between the true gradient and the ET/CET approximation w.r.t. return values in \textbf{A.}. \textbf{D.} Same as \textbf{C}, but for LunarLander-v2.
    }
    \label{fig:retro_rl}
\end{figure}

For the RL tasks, we observed similar trends. Increasing the number of CET states leads to improved performance on both CartPole and LunarLander (\cref{fig:retro_rl}A-B).
As before, the increase in the number of CET states also led to increased alignment with the true gradient (\cref{fig:retro_rl}C-D), although mostly in the second layer, which helps to explain the improved performance.

Altogether, our results demonstrate that when delays in credit signals are very long (on the order of minutes), and stacked (summing for each synaptic step), CETs can be used to store memory for previous activity in order to accurately estimate gradients and learn. As such, CETs would, in principle, permit credit assignment in situations where errors are propagated backwards via very slow chemical retrograde signals \citep{fitzsimonds1997propagation}. However, there is a depth limit beyond which the delay would be too large to accurately approximate the gradient.

\section{Discussion}\label{sec:discussion}

For organisms to learn, their brains must have mechanisms for handling delays between learning signals and past neural activity. 
Here we presented CETs, a generalization of classical eligibility traces and an abstract model of interacting biochemical processes within cells, as a candidate mechanism for bridging such delays. 
We showed that CETs enable learning over long delays on standard image datasets and RL settings, and found that by increasing the number of states in the eligibility trace cascade (with 1 state being equivalent to classical ETs) learning performance can be maintained with delays on the order of seconds.
We also explored how CETs contribute to the ongoing discussion around biologically plausible implementations of backpropagation \citep{fan2024contribute,liu2022biologically}. 
Here we tested the hypothesis that synaptic CETs and learning from salient examples enable slow cytoskeletal retroaxonal signals to carry gradient information recursively over layers -- an idea popular over two decades ago but since discarded \cite{harris2008stability}.
Again, we found CETs with a larger number of states improved performance. Though accumulating delays with network depth was still problematic for learning, CETs demonstrate that learning over timescales relevant for slow chemical signaling is feasible. 
As such our experiments with CETs validate retroaxonal signals a potential solution to the credit assignment problem. %
Overall, our work suggests that biochemical cascades within cells could provide a mechanism for generating eligibility traces that enable learning even when error signals arrive with a significant delay.    

Classical ETs are exponentially decaying ``memories'' of synaptic activation that are thought to be implemented by activation of biochemical processes, such as CaMKII activation or other 
protein kinases (such as PKA, PKC, ERK, MAPK), which are typically triggered by the activation of G-protein coupled receptors \citep{gerstner2018eligibility}. 
While the complexity of such interacting pathways has been recognized, there has been very little work exploring interactions between such biochemical processes for learning (though see \citep{friedrich2011spatio, huertas2016role}). %
In this context, we are building off work exploring how complex interactions between kinase cascades mediates plasticity \citep{zhang2021quantitative}.
More generally, CETs provide a normative explanation for the complexity of cascade interactions in the context of learning---the improved performance with higher order CETs could explain why cells use biochemical cascades rather than a single biochemical signal.

One of the well-known biological implausibilities of backpropagation is that it requires that weights in the forward pass be reused in the backward pass. 
In the context of biology, this algorithmic requirement is known as the weight transport problem \citep{grossberg1987competitive}.
Retroaxonal signals provide a potential solution to this problem because they pass back through the very same synapses used in the forward pass. 
As such, they could, in principle, carry information about the synaptic weights, thereby solving the weight transport problem \citep{fan2024contribute}.
However, the challenge with retroaxonal signals is that they are very slow, taking minutes to pass from the synapse to the cell body \citep{fitzsimonds1998retrograde}.
As we showed here, CETs provide a potential mechanism for making learning at such delays feasible.
Therefore, they open up the possibility of using retroaxonal signals for credit assignment. 
However, our results also showed that you can only stack such long delays over a few synaptic steps before learning deteriorates significantly, which would suggest that if retroaxonal signals are used for learning in the brain they would only be used for learning at relatively shallow ``depths''. 
Indeed this is consistent with experimental findings: for example \cite{hui2000selective, fitzsimonds1997propagation} only found retroaxonal potentiation and depression over one "layer" in cultured neurons (i.e. one synaptic step). 
Although it was originally suggested that the lack of further propagation may be due to the size of the plasticity change \cite{fitzsimonds1997propagation}, our results provide evidence that recursive propagation delays are problematic, even with CETs, and would require additional mechanisms such as direct reward signaling \citep{nokland2016direct}.

In summary, our work on CETs provides an extension to the classical ET approach for handling delays between activity and feedback error signals or rewards. 
We have demonstrated that cascades of biochemical processes could be used by cells to store more temporally precise memories of past cell activity. 
These memories could then be combined with delayed error signals to estimate loss gradients. 
Therefore, our work provides another potential means of understanding how the brain can learn complicated tasks in a biologically plausible manner.

\begin{ack}
This work was supported by: NSERC (Discovery Grant: RGPIN-2020-05105; Discovery Accelerator Supplement: RGPAS-2020-00031; Arthur B. McDonald Fellowship: 566355-2022); CIFAR (Canada AI Chair; Learning in Machine and Brains Fellowship); AccelNet (International network for brain-inspired computation, NSF 2019976). This research was enabled in part by support provided by (Calcul Québec) (\url{https://www.calculquebec.ca/en/}) and the Digital Research Alliance of Canada (\url{https://alliancecan.ca/en}). The authors acknowledge the material support of NVIDIA in the form of computational resources. 
The research was enabled in part by computational resources provided by Mila -- Quebec Artificial Intelligence Institute.

IA thanks Stephen Chung for his many helpful discussions and valuable comments on the reinforcement learning components of the project.

\end{ack}

\bibliographystyle{unsrtnat}
\bibliography{bibliography}

\clearpage
\appendix

\section{Cascading eligibility traces derivations}\label{sec:et_derivations}
\subsection{Update derivation}
The ET we presented in \cref{eq:ssm} has the form
\begin{equation*}
    \dot{\x}(t) = \bb A\,\x(t) + \bb b(t)\,,
\end{equation*}
which are solved by 
\begin{equation}
    \x(t) = \exp(\bb A t)\,\x(0) + \int_{0}^t \exp(\bb A (t - s))\,\bb b(s)\,ds\,.
    \label{app:eq:ode_solution}
\end{equation}
While the matrix exponent $\exp(\bb A t)$ is hard to compute in general, in our case,
\begin{equation}
\begin{pmatrix}
    \dot x^{1}_t\\
    \dots\\
    \dot x^{n-1}_t\\
    \dot x^{\mathrm{CET}}_t 
\end{pmatrix} = \begin{pmatrix}
    -\alpha & 0 & 0 & \dots & 0\\
    \dots & \dots & \dots & \dots & \dots\\
    0 & \dots & 1 & -\alpha & 0\\
    0 & \dots & 0 & 1 & -\alpha
\end{pmatrix} \begin{pmatrix}
    x^{1}_t\\
    \dots\\
    x^{n-1}_t\\
    x^{\mathrm{CET}}_t 
\end{pmatrix} + \begin{pmatrix}
    x_t\\
    \dots\\
    0\\
    0 
\end{pmatrix}\,,
\label{app:eq:ssm_matrix}
\end{equation}

therefore $\bb A = \alpha\, \bb I_n + \bb N$ for a nilpotent $\bb N$ (i.e. $\bb N^{n} = 0$), hence 
\begin{gather*}
    \exp((\alpha\, \bb I_n + \bb N)\,t)=\exp(\alpha\,t)\brackets{\sum_{i=0}^{n-1}\frac{1}{i!}N^i t^{i}}\\
    =\exp(\alpha\,t)\begin{bmatrix}
        1 & 0 & \dots & 0 & 0 \\
        t & 1 & 0 & \dots & 0 \\
        \frac{t^2}{2!} & t & 1 & \dots & 0\\
        \dots\\
        \frac{t^{n-1}}{(n-1)!} & \frac{t^{n-2}}{(n-2)!} & \dots & t &  1
    \end{bmatrix}\,.
\end{gather*}

Therefore, for $\bb x(0)=0$ and $\bb b(t)$ being non-zero only for the first coordinate, the last coordinate of $\x$ implements
\begin{equation*}
    x_{n}(t) = \int_0^t \exp(\alpha\,(t-s)) \frac{(t-s)^{n-1}}{(n-1)!}\,b_0(s)\,ds\,.
\end{equation*}

Moreover, if $\bb b(t)$ is a step-wise function taking on a new value every $\Delta t$ points, a single step of the integration between $t$ and $t+\Delta t$ can be computed (exactly) using \cref{app:eq:ode_solution} as 
\begin{gather*}
    \x(t+\Delta t) = \exp(\bb A \Delta t)\,\x(t) + \left[\int_{0}^{\Delta t} \exp(\bb A (\Delta t - s))\,\bb \,ds\right] \bb b(t)\\
    = \exp(\bb A \Delta t)\,\x(t) + \left[\int_{0}^{\Delta t} \exp(\bb A\,s)\,\bb \,ds\right] \bb b(t)\,.
\end{gather*}

As $\bb A$ is non-singular, we can integrate this solution further to obtain
\begin{equation}
    \x(t+\Delta t) = \exp(\bb A \Delta t)\,\x(t) + \left[\exp(\bb A\,\Delta t) - \bb I\right] \bb A^{-1} \bb b(t)\,.
    \label{app:eq:ssm_dt_update}
\end{equation}

\section{Implementation details}

\subsection{Alternative expression for stepwise inputs}

For experiments on visual tasks, we consider that a batch of inputs corresponds to a time-series where the batch dimension corresponds to the time dimension. In this case, the output of the state-space model in \cref{eq:ssm} can be obtained via a discrete convolution, which can be efficiently computed as a matrix multiplication. 

Starting from \cref{app:eq:ssm_dt_update}, we can denote $\bb M=\exp(\bb A \Delta t)$ and $\bb K=\left[\exp(\bb A\,\Delta t) - \bb I\right] \bb A^{-1}$, such that
\begin{equation*}
    \x(t+\Delta t) = \bb M\,\x(t) + \bb K\, \bb b(t)\,,
\end{equation*}
and therefore
\begin{equation*}
    \x(k\,\Delta t) = \bb M^k\,\x(0) + \sum_{i=0}^{k} \bb M^{k-i}\bb K\, \bb b(i\,\Delta t)\,.
\end{equation*}

If we assume the initial state was zero, $\x(0) = 0$, the SSM outputs $x^{\mathrm{CET}}_{k\,\Delta t}$ will be computed as (dropping the SSM superscript for convenience)

\begin{equation}
    \begin{bmatrix}
        x_{0} \\
        x_{\Delta t} \\
        \vdots \\
        x_{k\,\Delta t}
    \end{bmatrix} = \begin{bmatrix}
        g_{0} & 0 & \dots & 0 \\ 
        g_{1} & g_{0} & \dots & 0 \\ 
        \dots \\ 
        g_{k} & g_{k-1} & \dots & g_{0} \\ 
    \end{bmatrix} \begin{bmatrix}
        b_{0} \\
        b_{\Delta t} \\
        \vdots \\
        b_{k\,\Delta t}
    \end{bmatrix} = \bb G \bb b[t]
    \label{eq:mat_conv}
\end{equation}

where $g_{j} = (\bb M^{j}\bb K)_{n0}$.

Alternatively, to get a closed form expression for $g$, we may rewrite the stepwise constant input as the convolution of the appropriate impulse train with the rectangular function. Using $\delta$ for the Dirac delta and $\theta$ for the Heavyside function, we have
$$
    \begin{aligned}
        b(t) &= \sum_{i=1}^{\infty} a_i \delta(t-t_i), \; \textrm{rect}(t) = \theta(t) - \theta(t-1)\\ 
        \hat{b}(t) &= \sum_{i=1}^{\infty} a_i \textrm{rect}(t-t_i) = (b * \textrm{rect})(t)\\
        x(t) &= \hat{b} * g = b * (\textrm{rect} * g)\\
        &= \sum_{i=1}^{\infty} a_i (\textrm{rect} * g)(t - t_i).
    \end{aligned}
$$

We can therefore compute the exact continuous-time output with a discrete convolution using $\hat{g} = \textrm{rect} * g$.

\begin{equation}
    \begin{aligned}
        g(t) &= \theta(t) k t^n e^{-\frac{n}{T}t}\\
        \textrm{rect} * g &= \int_{-\infty}^{\infty} h(\tau) \textrm{rect} (t - \tau) d \tau\\
        &= k \int_{\max(0,t-1)}^{t} \tau^n e^{- \frac{n}{T} \tau} d \tau \\
        &= k ( \frac{T}{n})^{n+1} \left[ \gamma(n+1, \frac{n}{T}t) - \gamma (n+1, \frac{n}{T} \max(0,t-1))\right]
    \end{aligned}
    \label{app:eq:step_impulse_response}
\end{equation}

where $\gamma$ is the incomplete Gamma function. %

\subsection{Sparsification}\label{sec:mat_conv}

In matrix form, sparsification with indices in $\mathcal{T} = \{t_1,...,t_k\}$ then corresponds to a matrix multiplication with the diagonal matrix $\mathbf{S}_{\mathcal{T}} = \textrm{diag} \left( \mathbf{1}_{\mathcal{T}} \right)$, 

$$
    \mathbf{\tilde{x}[t]} = \mathbf{S}_{\mathcal{T}} \mathbf{x[t]}
$$

When using the Hebbian-like term $\bb h[t] = f'(\x[t]\T\w)\, \bb x[t]$ as inputs to the SSM, the gradient computation when both inputs and gradients are sparsified, respectively with $\mathcal{T},\mathcal{T}'$ will be

\begin{equation}
    \begin{aligned}
        \parderiv{L(\x[t]\T\w)}{\w} &= \mathbf{S}_{\mathcal{T}'} \bb \delta[t] \odot \bb G \mathbf{S_{\mathcal{T}}} \bb h[t]\\
        &= \mathbf{S}_{\mathcal{T}'} \bb \delta[t] \odot \mathbf{S}_{\mathcal{T}'} \bb G \mathbf{S}_{\mathcal{T}} \bb h[t].
    \end{aligned}
    \label{app:eq:conv_weight_updates}
\end{equation}

Since the sparsifying matrices are indicator functions, this is equivalent to indexing $\mathbf{\delta}_t$, $\mathbf{G}$, and $\mathbf{h[t]}$ at the appropriate positions defined by $\mathcal{T}$ and $\mathcal{T}'$. When simulating sparsity, we obtain the original input presentation indices $\mathcal{T}$ as the salient image indices and compute the arrival time gradient indices as $\mathcal{T}' = \mathcal{T} + (m-1)T$. The gradient over the batch is then computed by summing over the time—or equivalently, batch—dimension.

\section{Experimental details}\label{app:sec:experiments}

\paragraph{Visual experiments.} For visual experiments, we consider the batch dimension to be the time dimension, and we compute the delayed signals over the batch dimension using a matrix convolution as described in \cref{sec:mat_conv}. The experiments in \cref{subseq:btsp} use a batch size of $128$ samples, while the experiments in \cref{subseq:retro} use a batch size of $1280$ samples, where only the samples with the top $1.25\%$ of training losses are used. Networks were trained using the cross-entropy loss and the AdamW optimizer with \(\beta_1=0.9,\, \beta_2=0.999\). The learning rate was scaled with a linear warm-up over $10\%$ and $20\%$ of the training steps for \cref{subseq:btsp} and \cref{subseq:retro}, respectively, followed by cosine annealing to $10\%$ of the initial learning rate. For the experiments in \cref{subseq:btsp}, the maximal learning rate was selected from a logarithmic grid of 5 points spanning \(10^{-3}\) to \(10^{-7}\), and the weight decay was chosen from the set \(\{0.1,\, 0.01,\, 0.001,\, 0.0\}\). For the CIFAR-10 experiments in \cref{subseq:retro}, the maximal learning rate was selected from $\{5\times10^{-5},\, 2.5\times10^{-5},\, 1\times10^{-5},\, 7.5\times10^{-6},\, 5\times10^{-6},\, 2.5\times10^{-6},\, 1\times10^{-6}\}$, and the weight decay was fixed to $0.1$. Hyper-parameters for the MNIST experiments in \cref{subseq:retro}, were searched the same way as for \cref{subseq:btsp}. Hyper-parameters presented in \cref{tab:btsp_cifar_10_config_vis,tab:btsp_mnist_config_vis,tab:retro_cifar_configs_vis,tab:retro_mnist_config_vis} were independently selected using a 90\%/10\% split of the standard training set, and the models were retrained using the standard training set and tested on the standard test set. All experiments in \cref{subseq:btsp} as well as the MNIST experiments in \cref{subseq:retro} were run for $20000$ training steps. The CIFAR-10 experiments in \cref{subseq:retro} were run for $1562500$ steps, in large part due to lower learning rates. Data augmentation using random horizontal flips was applied only to the CIFAR-10 experiments.

\begin{table}[ht]
    \centering
    \small %
    \setlength{\tabcolsep}{4pt} %
    \begin{minipage}{0.48\textwidth}
        \centering
        \caption{Experiment configurations for CIFAR-10 experiments at behavioural timescales.}
        \label{tab:btsp_cifar_10_config_vis}
        \begin{tabular}{rrrr}
        \toprule
        \textbf{CET order} & \textbf{delay} & \textbf{lr} & \textbf{weight decay} \\
        \midrule
        1  &  0.2  & 1e-3   & 1e-1   \\
        2  &  0.2  & 1e-3   & 1e-1   \\
        6  &  0.2  & 1e-3   & 1e-1   \\
        10 &  0.2  & 1e-3   & 1e-1   \\
        1  &  0.6  & 1e-3   & 1e-3   \\
        2  &  0.6  & 1e-3   & 1e-1   \\
        6  &  0.6  & 1e-3   & 1e-1   \\
        10 &  0.6  & 1e-3   & 1e-1   \\
        1  &  1.0  & 1e-4   & 0      \\
        2  &  1.0  & 1e-3   & 1e-2   \\
        6  &  1.0  & 1e-3   & 1e-1   \\
        10 &  1.0  & 1e-3   & 1e-1   \\
        1  &  2.0  & 1e-4   & 1e-3   \\
        2  &  2.0  & 1e-4   & 1e-3   \\
        6  &  2.0  & 1e-3   & 1e-3   \\
        10 &  2.0  & 1e-3   & 0      \\
        1  &  4.0  & 1e-4   & 1e-3   \\
        2  &  4.0  & 1e-4   & 0      \\
        6  &  4.0  & 1e-4   & 1e-3   \\
        10 &  4.0  & 1e-4   & 1e-2   \\
        1  & 10.0  & 1e-4   & 0      \\
        2  & 10.0  & 1e-4   & 1e-2   \\
        6  & 10.0  & 1e-4   & 1e-3   \\
        10 & 10.0  & 1e-4   & 1e-3   \\
        \bottomrule
        \end{tabular}
    \end{minipage}
    \hfill
    \begin{minipage}{0.48\textwidth}
        \centering
        \caption{Experiment configurations for MNIST experiments at behavioural timescales.}
        \label{tab:btsp_mnist_config_vis}
        \begin{tabular}{rrrr}
        \toprule
        \textbf{CET order} & \textbf{delay} & \textbf{lr} & \textbf{weight decay} \\
        \midrule
        1  &  0.2  & 1e-3   & 1e-1   \\
        2  &  0.2  & 1e-3   & 1e-3   \\
        6  &  0.2  & 1e-3   & 1e-2   \\
        10 &  0.2  & 1e-3   & 1e-2   \\
        1  &  0.6  & 1e-3   & 1e-3   \\
        2  &  0.6  & 1e-3   & 1e-1   \\
        6  &  0.6  & 1e-3   & 1e-1   \\
        10 &  0.6  & 1e-3   & 1e-3   \\
        1  &  1.0  & 1e-3   & 0      \\
        2  &  1.0  & 1e-3   & 0      \\
        6  &  1.0  & 1e-3   & 0      \\
        10 &  1.0  & 1e-3   & 1e-3   \\
        1  &  2.0  & 1e-3   & 1e-3   \\
        2  &  2.0  & 1e-3   & 1e-2   \\
        6  &  2.0  & 1e-3   & 0      \\
        10 &  2.0  & 1e-3   & 1e-2   \\
        1  &  4.0  & 1e-3   & 0      \\
        2  &  4.0  & 1e-3   & 0      \\
        6  &  4.0  & 1e-3   & 0      \\
        10 &  4.0  & 1e-3   & 1e-3   \\
        1  & 10.0  & 1e-4   & 0      \\
        2  & 10.0  & 1e-4   & 1e-2   \\
        6  & 10.0  & 1e-3   & 0      \\
        10 & 10.0  & 1e-3   & 1e-3   \\
        \bottomrule
        \end{tabular}
    \end{minipage}
\end{table}

\begin{table}[ht]
    \centering
    \small
    \setlength{\tabcolsep}{6pt} %
    \begin{minipage}
    {0.48\textwidth}
        \centering
        \caption{Experiment configurations for CIFAR-10 experiments at retrograde timescales.}
        \label{tab:retro_cifar_configs_vis}
        \begin{tabular}{rrr}
        \toprule
        \textbf{CET order} & \textbf{lr} & \textbf{weight decay} \\
        \midrule
        1  & 1e-5  & 1e-1 \\
        2  & 1e-5  & 1e-1 \\
        6  & 5e-5  & 1e-1 \\
        10 & 5e-5  & 1e-1 \\
        \bottomrule
        \end{tabular}
    \end{minipage}\hfill
    \begin{minipage}{0.48\textwidth}
    \centering
    \caption{Configurations for MNIST experiments at retrograde timescales.}
    \label{tab:retro_mnist_config_vis}
    \begin{tabular}{rrr}
    \toprule
    \textbf{CET order} & \textbf{lr} & \textbf{weight decay} \\
    \midrule
    1  & 1e-4 & 1e-3  \\
    2  & 1e-4 & 1e-1  \\
    6  & 1e-4 & 1e-3  \\
    10 & 1e-4 & 0.0   \\
    \bottomrule
    \end{tabular}
    \end{minipage}
\end{table}

\paragraph{RL.} In Actor-Critic, we train the Critic using the standard $\lambda$-return, while the Actor is trained using RL eligibility traces (see Algorithm~\ref{algo:et_rl}). The term $\nabla_{\theta} \log \pi_{\theta}(a_t \mid s_t)$ in Algorithm~\ref{algo:et_rl} refers either to the true gradient obtained via backpropagation or to its ET/CET approximations, computed as the product of the top-level gradient signal and the ET/CET output. The CET update is computed using Eq.~\ref{app:eq:ssm_dt_update}.

\begin{algorithm}
\caption{Actor learning via RL eligibility traces.}
\begin{algorithmic}[1]
\STATE Initialize actor parameters $\theta$, RL eligibility trace vector $e = 0$, gradient accumulator $\nabla_{\theta} L = 0$, learning rate $\alpha$, and trace decay $\lambda$.
\STATE Sample initial state $s_0$ from the environment
\FOR{$t \in 0, \ldots, L$}
  \STATE Select action $a_t \sim \pi_{\theta}(\cdot \mid s_t)$
  \STATE Take action $a_t$, observe $r_t, s_{t+1}$
  \STATE Update RL eligibility trace vector: $e = \lambda \gamma e + \nabla_{\theta} \log \pi_{\theta}(a_t \mid s_t)$
  \STATE Compute TD error: $\eta_t = r_t + \gamma V(s_{t+1}) - V(s_t)$
  \STATE Accumulate gradient: $\nabla_{\theta} L = \nabla_{\theta} L + \eta_t e$
  \IF{$t \mod n = 0$}
    \STATE Update actor: $\theta = \theta + \alpha \nabla_{\theta} L$
    \STATE Reset gradient: $\nabla_{\theta} L = 0$
  \ENDIF
\ENDFOR
\end{algorithmic}
\label{algo:et_rl}
\end{algorithm}

The learning rate was selected from the set ${2.5\mathrm{e}{-4},\ 5\mathrm{e}{-4},\ 9\mathrm{e}{-4},\ 1\mathrm{e}{-4}}$ based on performance for all experiments. For all classic ET runs, the ET discount factor, $\beta$, was chosen from ${0.5,\ 0.7,\ 0.9,\ 0.99}$. Additionally, we used two normalization schemes for CET outputs: \textit{area} and \textit{peak} normalizations. In \textit{area} normalization, the CET output is scaled so that the response to a unit input integrates to one across all future states. In \textit{peak} normalization, the CET output is scaled such that the maximum response to a unit input is one. For Fig.\ref{fig:retro_rl}, this normalization hyperparameter was also searched. 

For the Critic, we used either the same architecture as the Actor, a three-layer MLP with hidden dimension 256, or a convolutional neural network (CNN) for MinAtar/SpaceInvaders-v0, consisting of three convolutional layers (kernel size 3, zero-padding 1) followed by two fully connected layers. The ReLU activation function was used in all experiments. For CartPole and LunarLander,  we also controlled the simulated time elapsed between the environment receiving an action from the agent and producing the corresponding next state and reward. This time was set to 200 ms for behavioral timescale experiments (Section~\ref{subseq:btsp}) and 300 ms for retrograde signaling (Section~\ref{subseq:retro}), based on the time modeling assumptions described in the referenced sections.

To better preserve gradient alignment with ET/CET in the first layer, we ensured positive inputs by doubling the input dimensionality and representing each original dimension with separate positive and negative components.

The remaining hyperparameters used in the experiments are summarized in Table~\ref{table:rl_hp}, and tuned hyperparameters are reported in Tables~\ref{table:rl_btsp_tuned_hp} and \ref{table:rl_retro_tuned_hp}. Note that for MinAtar/SpaceInvaders, we use a randomly sampled $\lambda$ value, as we found this improves performance in this environment. A separate $\lambda$ is sampled independently for every learned parameter, which is feasible due to our RL eligibility traces implementation of Actor learning.

\begin{table}
	\caption[Actor-Critic hyperparameters used in RL experiments]{Hyperparameters used in RL experiments.}
	\label{table:rl_hp}
	\begin{center}	
		\begin{tabular}{lllll}
			\toprule[0.1ex]
			\textbf{Parameter} & \textbf{Value} \\
                \\
                \textbf{Common} \\
                Optimizer & Adam \\                
			Adam beta & (0.9, 0.999) \\
                Adam epsilon & 1e-5 \\ 
                Weight decay & 0 \\
                Policy entropy regularization coefficient & 0.01 \\ 
                Maximum gradient norm for clipping & 0.5 \\
                Learning rate & Tuned \\
                Discount rate $\gamma$ & 0.99 \\
			\\
                \textbf{CartPole-v1} \\
                Total number of samples & $5\_000\_000$ \\
                Number of environments & 4 \\
                Number of steps to accumulate a policy gradient & 128 \\
                Lambda for general advantage estimation & 0.95 \\
                Anneal lr & True \\    
                CET normalization & Peak \\
                \\
			\textbf{LunaLander-v2} \\
                Total number of samples & $5\_000\_000$ \\
                Number of environments & 4 \\
                Number of steps to accumulate a policy gradient & 128 \\
                Lambda for general advantage estimation & 0.95 \\
                Anneal lr & False \\     
                CET normalization & Area or Tuned \\
                \\
                \textbf{MinAtar/SpaceInvaders-v0} \\
                Total number of samples & $10\_000\_000$ \\
                Number of environments & 32 \\
                Number of steps to accumulate a policy gradient & 32 \\
                Lambda for general advantage estimation & Random Uniform(0.1, 0.99) \\
                Anneal lr & False \\     	
                CET normalization & Peak \\
                \bottomrule[0.25ex]
		\end{tabular}
	\end{center}
\end{table}

\begin{table}[ht]
    \centering
     \caption{Optimal learning rate and ET discounting factor configurations, $\beta$, for experiments at behavioral timescales.}
     \label{table:rl_btsp_tuned_hp}
    \scriptsize %
    \setlength{\tabcolsep}{3pt} %
    \begin{minipage}{0.3\textwidth}
        \centering
    \begin{tabular}{rrrl}
    \multicolumn{4}{c}{\textbf{CartPole-v1}} \\  
    \toprule
 \textbf{CET order} & \textbf{delay} & \textbf{lr} & \textbf{$\beta$} \\
    \midrule
          backprop &          - & 0.00090 &   - \\
            1 &          2 & 0.00050 &   0.5 \\
            2 &          2 & 0.00025 &     - \\
            5 &          2 & 0.00025 &     - \\
            8 &          2 & 0.00025 &     - \\
           10 &          2 & 0.00025 &     - \\
            1 &          4 & 0.00050 &   0.9 \\
            2 &          4 & 0.00025 &     - \\
            8 &          4 & 0.00025 &     - \\
            5 &          4 & 0.00025 &     - \\
           10 &          4 & 0.00050 &     - \\
            1 &          8 & 0.00050 &  0.99 \\
            2 &          8 & 0.00025 &     - \\
            5 &          8 & 0.00025 &     - \\
            8 &          8 & 0.00025 &     - \\
           10 &          8 & 0.00050 &     - \\
            1 &         32 & 0.00010 &  0.99 \\
            2 &         32 & 0.00025 &     - \\
            5 &         32 & 0.00025 &     - \\
            8 &         32 & 0.00025 &     - \\
           10 &         32 & 0.00010 &     - \\
            1 &         64 & 0.00010 &  0.99 \\
            2 &         64 & 0.00025 &     - \\
            5 &         64 & 0.00090 &     - \\
            8 &         64 & 0.00090 &     - \\
           10 &         64 & 0.00050 &     - \\
            1 &        128 & 0.00050 &   0.5 \\
            2 &        128 & 0.00010 &     - \\
            5 &        128 & 0.00025 &     - \\
            8 &        128 & 0.00025 &     - \\
           10 &        128 & 0.00025 &     - \\
    \bottomrule
    \end{tabular}
    \end{minipage}
    \hfill
    \begin{minipage}{0.3\textwidth}
        \centering
        \begin{tabular}{rrrr}
        \multicolumn{4}{c}{\textbf{LunarLander-v2}} \\  
        \toprule
        \textbf{CET order} & \textbf{delay} & \textbf{lr} & \textbf{$\beta$} \\
        \midrule
      backprop &          - & 0.00050 &     - \\
        1 &          2 & 0.00090 &  0.99 \\
        2 &          2 & 0.00050 &     - \\
        5 &          2 & 0.00050 &     - \\
        8 &          2 & 0.00050 &     - \\
       10 &          2 & 0.00050 &     - \\
        1 &          4 & 0.00090 &  0.99 \\
        2 &          4 & 0.00050 &     - \\
        5 &          4 & 0.00050 &     - \\
        8 &          4 & 0.00050 &     - \\
       10 &          4 & 0.00090 &     - \\
        1 &          8 & 0.00090 &  0.99 \\
        2 &          8 & 0.00050 &     - \\
        5 &          8 & 0.00050 &     - \\
        8 &          8 & 0.00090 &     - \\
       10 &          8 & 0.00050 &     - \\
        1 &         32 & 0.00090 &  0.99 \\
        2 &         32 & 0.00090 &     - \\
        5 &         32 & 0.00090 &     - \\
        8 &         32 & 0.00050 &     - \\
       10 &         32 & 0.00025 &     - \\
        1 &         64 & 0.00090 &  0.99 \\
        2 &         64 & 0.00090 &     - \\
        5 &         64 & 0.00090 &     - \\
        8 &         64 & 0.00090 &     - \\
       10 &         64 & 0.00050 &     - \\
        1 &        128 & 0.00050 &   0.7 \\
        2 &        128 & 0.00090 &     - \\
        5 &        128 & 0.00090 &     - \\
        8 &        128 & 0.00090 &     - \\
       10 &        128 & 0.00090 &     - \\
        \bottomrule
        \end{tabular}
    \end{minipage}
    \hfill
    \begin{minipage}{0.31\textwidth}
    \centering
    \begin{tabular}{rrrr}
    \multicolumn{4}{c}{\textbf{MinAtar/SpaceInvaders-v0}} \\  
    \toprule
    \textbf{CET order} & \textbf{delay} & \textbf{lr} & \textbf{$\beta$} \\
    \midrule
    backprop &          1 & 0.00090 &     - \\
        1 &          1.2 & 0.00025 &   0.9 \\
        2 &          1.2 & 0.00090 &     - \\
        5 &          1.2 & 0.00050 &     - \\
        8 &          1.2 & 0.00050 &     - \\
       10 &          1.2 & 0.00090 &     - \\
        1 &          2 & 0.00090 &  0.99 \\
        2 &          2 & 0.00050 &     - \\
        5 &          2 & 0.00050 &     - \\
        8 &          2 & 0.00050 &     - \\
       10 &          2 & 0.00050 &     - \\
        1 &          4 & 0.00090 &  0.99 \\
        2 &          4 & 0.00050 &     - \\
        5 &          4 & 0.00090 &     - \\
        8 &          4 & 0.00050 &     - \\
       10 &          4 & 0.00050 &     - \\
        1 &          8 & 0.00050 &  0.99 \\
        2 &          8 & 0.00050 &     - \\
        5 &          8 & 0.00090 &     - \\
        8 &          8 & 0.00025 &     - \\
       10 &          8 & 0.00050 &     - \\
    \bottomrule
    \end{tabular}
    \end{minipage}

\end{table}

\begin{table}[ht]
    \centering
    \caption{Optimal learning rate, CET normalization, and ET discounting factor configurations, $\beta$, for experiments at retrograde timescales. }
    \label{table:rl_retro_tuned_hp}
    \small
    \setlength{\tabcolsep}{6pt} %
    \begin{minipage}{0.48\textwidth}
        \centering
    \begin{tabular}{rrrr}
    \multicolumn{4}{c}{\textbf{CartPole-v1}} \\  
    \toprule
     \textbf{CET order}  &  \textbf{lr} &  \textbf{normalization} & $\beta$ \\
    \midrule
             1 & 0.00010 &      - & 0.9 \\
             2 & 0.00010 &       peak & - \\
             5 & 0.00010 &       peak & - \\
            10 & 0.00025 &       peak & - \\
    \bottomrule
    \end{tabular}
        \label{tab:retro_cifar_configs}
    \end{minipage}\hfill
    \begin{minipage}{0.48\textwidth}
    \centering
    \begin{tabular}{rrrr}
    \multicolumn{4}{c}{\textbf{LunarLander-v2}} \\  
    \toprule
    \textbf{CET order}  &  \textbf{lr} &  \textbf{normalization} & $\beta$ \\
    \midrule
         1 & 0.00025 &      -  & 0.5 \\
         2 & 0.00050 &       area & - \\
         5 & 0.00025 &       area & - \\
        10 & 0.00050 &       peak & - \\
    \bottomrule
    \end{tabular}
    \label{tab:retro_mnist_config}
    \end{minipage}
\end{table}

\paragraph{Compute.} All experiments were done on RTX 8000 and A100 GPUs. Each MNIST run takes between 3 and 10 minutes on an RTX 8000 GPU, while each CIFAR-10 run takes approximately 30 minutes for \cref{subseq:btsp} and up to 24 hours for \cref{subseq:retro} on a RTX 8000. Each RL run takes approximately 1-2 hours to complete.

\section{Supplemental Results}\label{app:sec:supp_results}

\begin{figure}[h!]
    \centering
\includegraphics{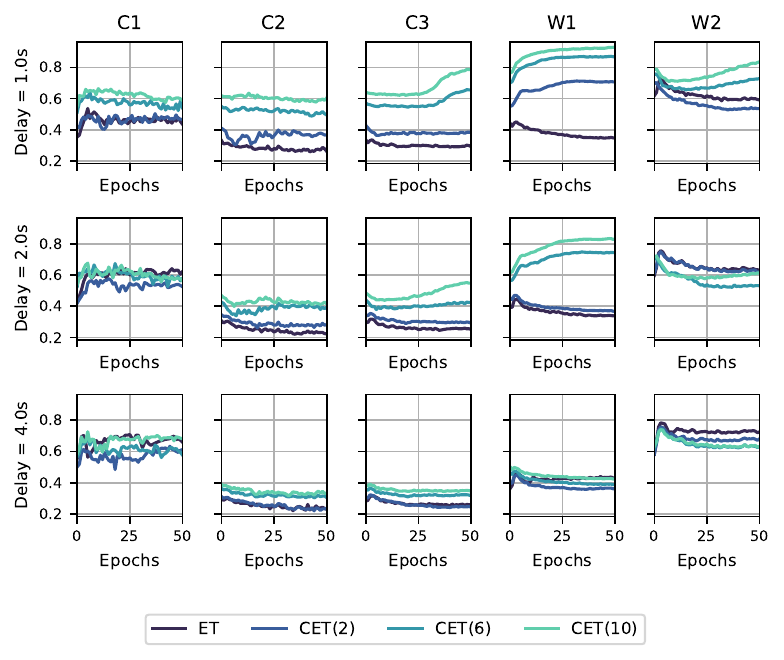}
\caption{Cosine similarity for different layers of a CNN between between true gradients and ETs or CETs approximated gradients for all considered environments during training on CIFAR-10. C1-3: convolutional layers 1-3; W1-2: MLP layers 1-2.}
\label{supp:fig:btsp_full_visual_sim}
\end{figure}

\begin{figure}[h!]
    \centering
\includegraphics[width=0.9\textwidth]{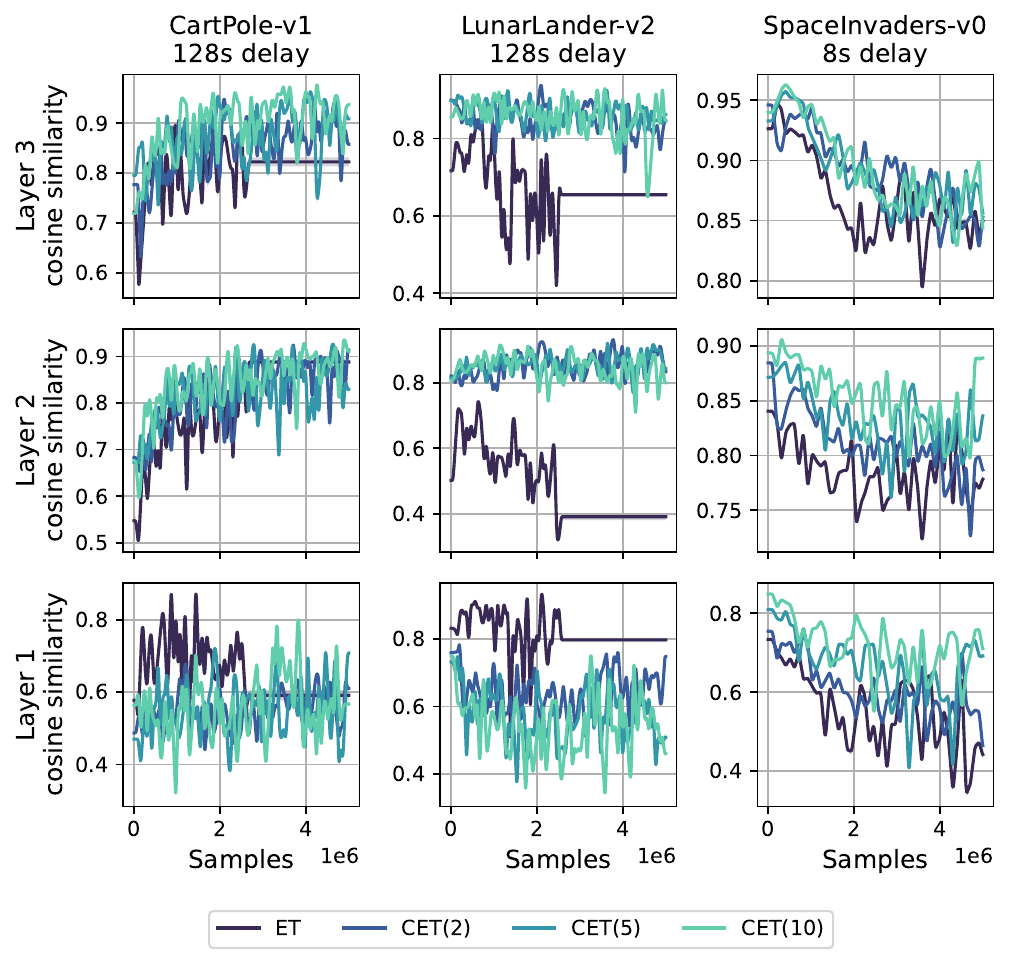}
\caption{Average cosine similarity between the true gradients and those approximated by ETs or CETs during training, computed for each layer of a 3-layer MLP across all considered environments. The delay is set to the maximum behavioral-timescale value reported in the main text.}
\label{supp:fig:btsp_rl_sim}
\end{figure}

To complement the analyses in the main text, we provide additional results on gradient alignment across all layers for different tasks. First, we show in \cref{supp:fig:btsp_full_visual_sim} that the separation observed in \cref{fig:btsp_sim} generally holds across different layers. A similar trend is observed for RL tasks in Fig.~\ref{supp:fig:btsp_rl_sim}, although the separation is much noisier and sometimes does not hold for the first layer. We hypothesize that the training dynamics of ET and CET can differ significantly, guiding parameters to distinct regions in the loss landscape. In these regions, gradient alignment might occasionally be higher for ET, yet overall performance remains lower. 

\begin{figure}[h!]
    \centering
\includegraphics{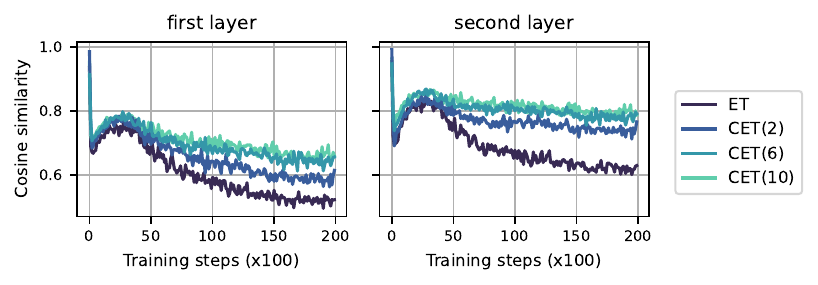}
\caption{Cosine similarity between the true gradient and the ET/CET approximation in a retrograde setting. Each plot shows an individual layer of an MLP during training on MNIST}
\label{supp:fig:retro_mnist_sim_overtime}
\end{figure}

\begin{figure}[h!]
    \centering
\includegraphics{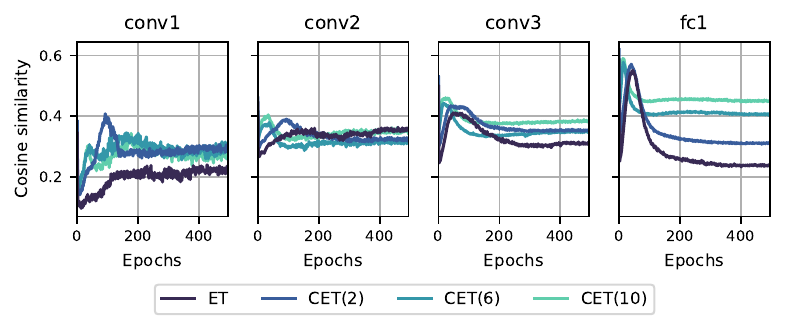}
\caption{Cosine similarity between the true gradient and the ET/CET approximation in a retrograde setting. Each plot shows an individual layer of a CNN during training on CIFAR-10}
\label{supp:fig:retro_cifar_sim_overtime}
\end{figure}

We can also see in \cref{supp:fig:retro_mnist_sim_overtime} that the separation between the different CETs is still evident when plotting the similarity against training steps for MNIST. However, \cref{supp:fig:retro_cifar_sim_overtime} shows that this relationship is muddied for CIFAR-10, which justifies plotting against accuracy.

\begin{figure}[h!]
    \centering
        \begin{minipage}[t]{0.33\textwidth}
\centering
\includegraphics[width=0.99\textwidth]{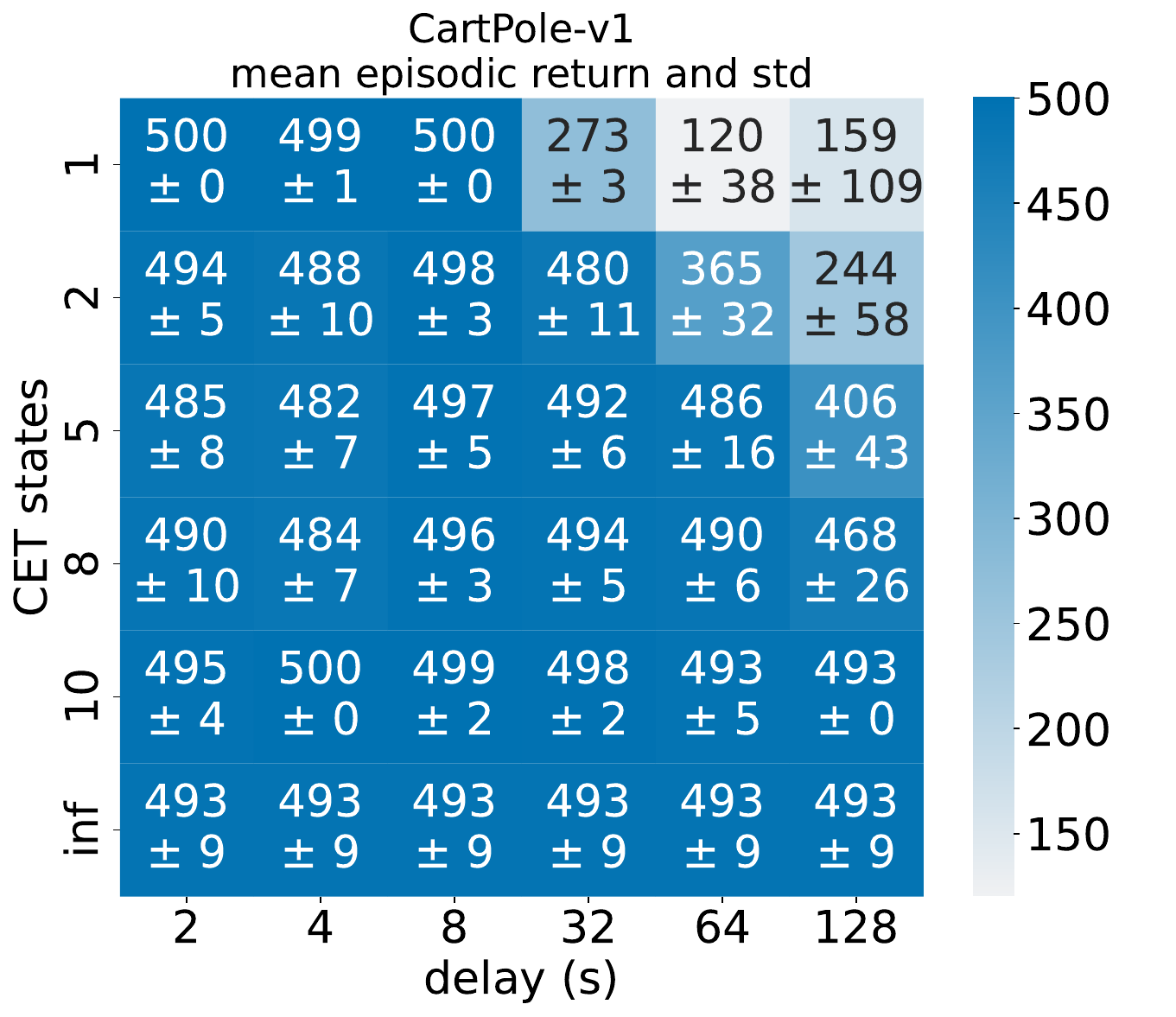}
    \end{minipage}\hfill
    \begin{minipage}[t]{0.33\textwidth}
\centering
        \includegraphics[width=0.99\textwidth]{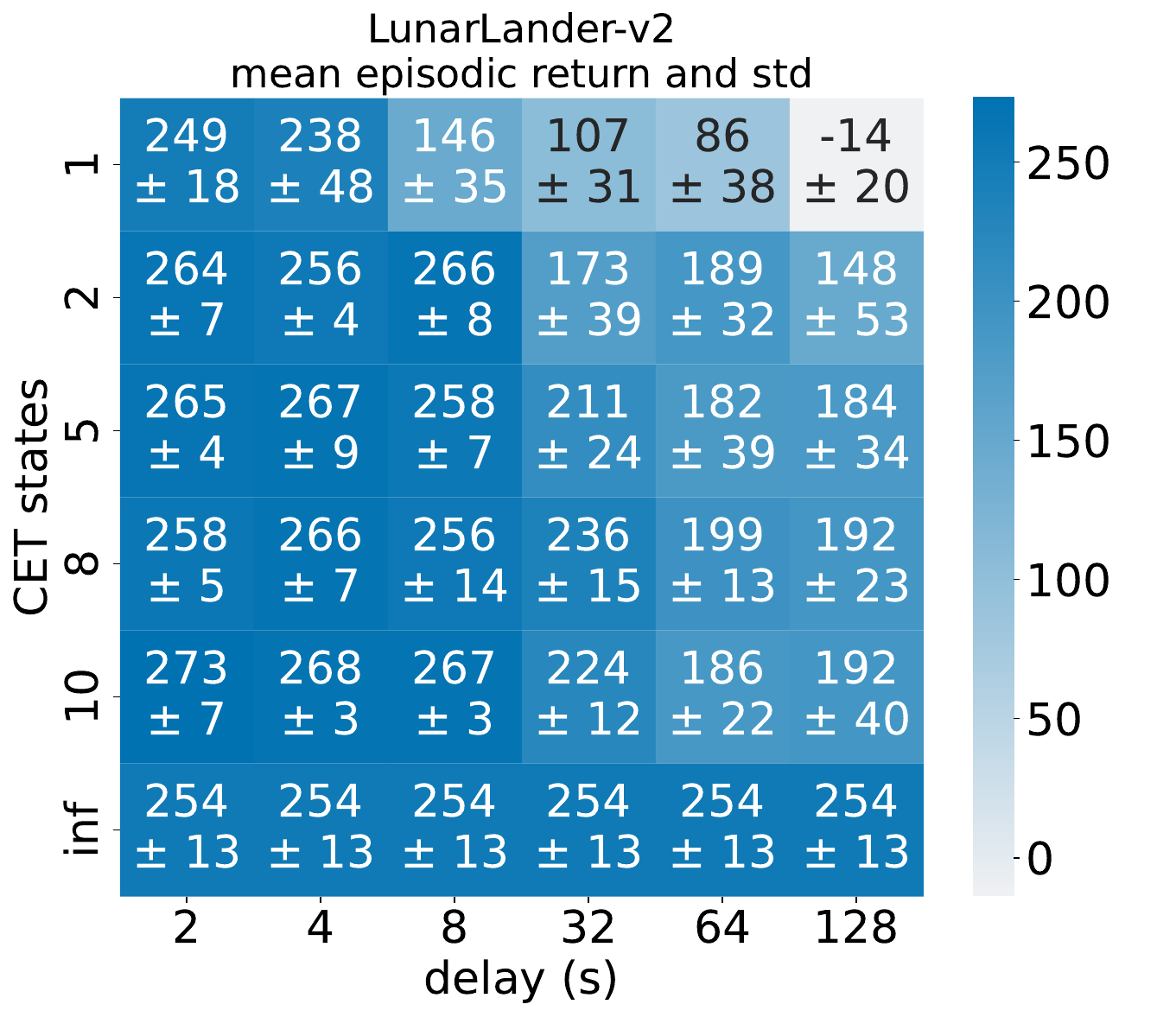}
    \end{minipage}\hfill
\begin{minipage}[t]{0.33\textwidth}
\centering
        \includegraphics[width=0.99\textwidth]{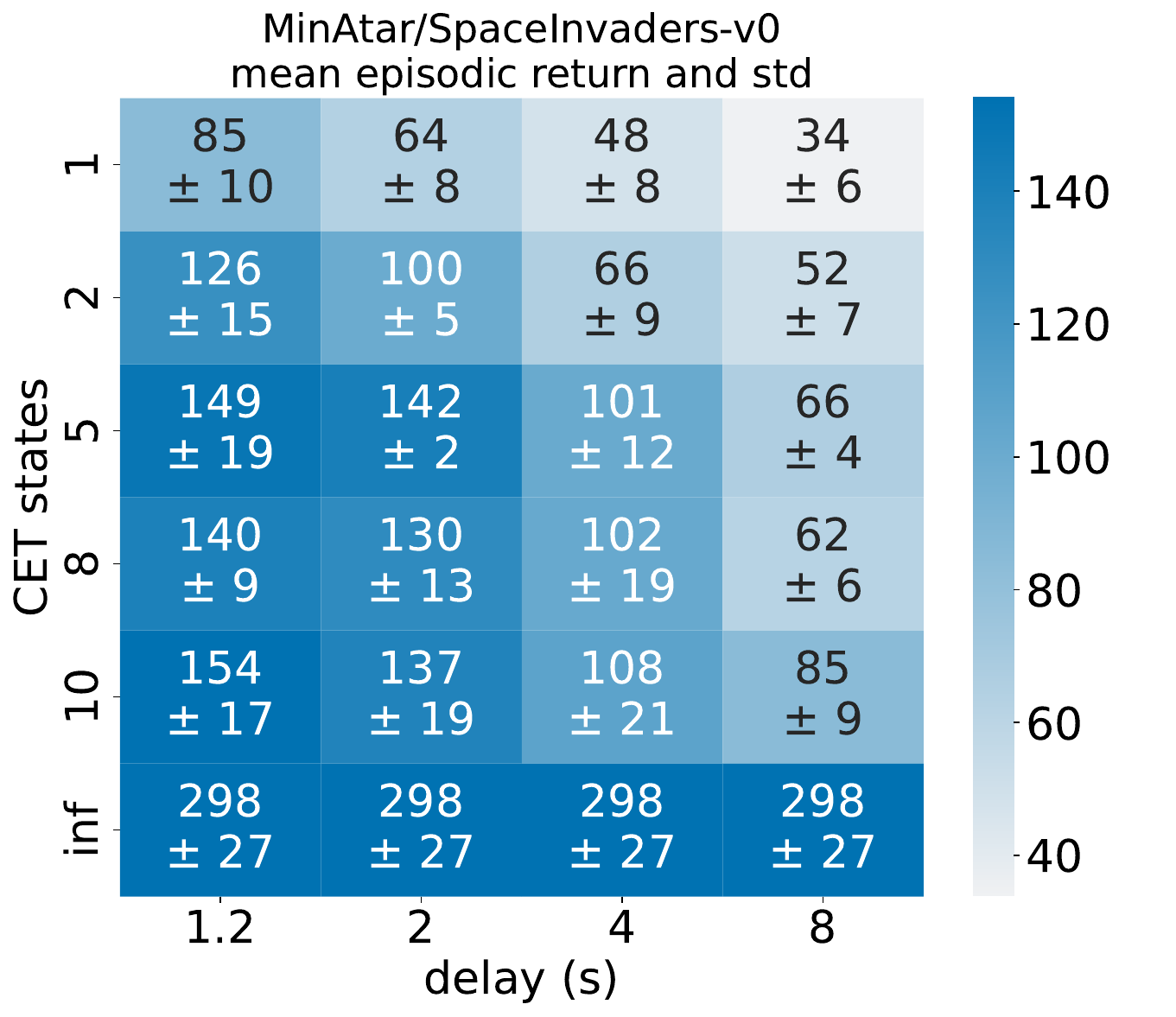}
    \end{minipage}\hfill

\caption{Same as Fig.~\ref{fig:btsp_rl}, but with standard deviation over three seeds. ``inf'' refers to backpropagation baseline.  
}
\label{fig:btsp_std_rl}
\end{figure}

We also report results with standard deviations for RL tasks at the behavioral timescale in Fig.~\ref{fig:btsp_std_rl} to show reliability.

Additionally, we note that no experiments were conducted with MinAtar/SpaceInvaders at the retrograde timescale. As shown in the rightmost heatmap of Fig.~\ref{fig:btsp_rl}, CET does not scale well to longer timescales on SpaceInvaders, exhibiting only modest performance with an 8-second delay and consequently failing at a 120-second delay at the retrograde timescale (not shown).

\end{document}